\documentclass[a4paper,fleqn]{cas-dc}

\usepackage[numbers]{natbib}
\usepackage{subfig}
\usepackage{upgreek}
\usepackage{siunitx}

\newcommand{\neqcm}{n$_{\mathrm{eq}}$/cm$^{2}$\xspace}
\def\vbulk {V$_{\mathrm{bulk}}$\xspace}
\def\vgain{V$_{\mathrm{gain}}$\xspace}
\def\deltavbulk{$\Delta$V$_{\mathrm{bulk}}$\xspace}

\def\drift     {\raisebox{.5pt}{\textcircled{\raisebox{-.9pt} {1}}}\xspace}
\def\induction {\raisebox{.5pt}{\textcircled{\raisebox{-.9pt} {2}}}\xspace}

\def\siemfive{SiEM \SI{5}{\micro\meter}\xspace}
\def\siemten{SiEM \SI{10}{\micro\meter}\xspace}

\DeclareSIUnit{\sample}{S}

\begin{document}

\let\WriteBookmarks\relax
\renewcommand{\floatpagefraction}{0.7}
\renewcommand{\textfraction}{0.1}

\shorttitle{Characterisation of the first SiEM demonstrators}
\shortauthors{De Benedetti et al.}
\title[mode=title]{Detection of minimum ionising particles with internal gain in the first Silicon Electron Multiplier demonstrators}

\author[1,2]{Federico {De Benedetti}}[type=author,auid=0000-0002-7960-3116]
\cormark[1]
\ead{federico.de.benedetti@cern.ch}
\author[1]{Sara Cesare}[type=author,auid=0000-0003-0886-7111]
\author[1]{Victor Coco}[type=author,auid=0000-0002-5310-6808]
\author[1]{Paula Collins}[type=author,auid=0000-0003-1437-4022]
\author[1]{Raphael Dumps}[type=author,auid=0000-0001-5524-0085]
\author[1]{Edgar Cid Lemos}[type=author,auid=0000-0003-3001-6268]
\author[1,3,4]{Samuel Mico}[type=author,auid=0009-0003-7101-8144]
\author[1]{Janina Nicolini}[type=author,auid=0000-0001-9034-3637]
\author[1,5]{Alfonso Puicercus Gomez}[type=author,auid=0009-0005-9982-6383]
\author[1]{Efren Rodriguez Rodriguez}[type=author,auid=0000-0002-7973-8061]
\author[1,6]{Lotte Sikkema}[type=author,auid=0009-0001-1366-9059]
\author[7]{Hironori Sonobe}[type=author,auid=0009-0009-2885-5880]
\author[2]{Abraham Gallas Torreira}[type=author,auid=0000-0002-2745-7954]
\author[1]{Morag Williams}[type=author,auid=0000-0001-7765-8941]

\affiliation[1]{organization={CERN}, city={Geneva}, postcode={1217}, country={Switzerland}}
\affiliation[2]{organization={Instituto Galego de Física de Altas Enerxías (IGFAE) and Universidade de Santiago de Compostela (USC)}, city={Santiago de Compostela}, postcode={15705}, country={Spain}}
\affiliation[3]{organization={Institute of Physics of the Czech Academy of Sciences}, city={Prague}, postcode={18200}, country={Czech Republic}}
\affiliation[4]{organization={Czech Technical University in Prague, Faculty of Nuclear Sciences and Physical Engineering}, city={Prague}, postcode={11519}, country={Czech Republic}}
\affiliation[5]{organization={University of Groningen, Faculty of Science and Engineering}, city={Groningen}, postcode={9747 AG}, country={The Netherlands}}
\affiliation[6]{organization={University of Amsterdam, Faculty of Sciences}, city={Amsterdam}, postcode={1012WX}, country={The Netherlands}}
\affiliation[7]{organization={Hamamatsu Photonics Kabushiki Kaisha, Solid State Division},
city={Hamamatsu, Shizuoka},postcode={435-8558},country={Japan}}

\cortext[1]{Corresponding author}

\begin{abstract}
The Silicon Electron Multiplier (SiEM) is a novel sensor concept for minimum ionising particle (MIP) detection that uses internal gain achieved through a composite electrode structure embedded within the silicon bulk. Unlike conventional gain sensors (LGADs, APDs), the SiEM achieves amplification through applied electric potentials rather than doped gain layers, potentially offering superior radiation hardness above 10$^{16}$\neqcm. Following initial conceptual studies and simulation work, several fabrication approaches were explored, including Deep Reactive Ion Etching (DRIE) at Centro Nacional de Microelectrónica (CNM) and Metal Assisted Chemical Etching (MACEtch) at the Paul Scherrer Institute (PSI). This paper reports the first experimental results from functional SiEM demonstrators produced by Hamamatsu Photonics K. K. (HPK) using a single-electrode configuration with modified geometry. Electrical characterisation through I-V measurements and test beam studies with minimum ionising particles demonstrate clear evidence of charge multiplication, with gains comparable to simulation predictions.
\end{abstract}


\maketitle

\section{Introduction}\label{sec:introduction}
 
\subsection{Motivation and Challenges}
Future particle physics experiments require silicon sensors with increasingly demanding specifications: time resolution below \SI{50}{\pico\second}, spatial resolution below \SI{10}{\micro\meter}, radiation tolerance above 10$^{16}$\neqcm, and the capability to operate at high particle rates~\cite{cern-rd}. Current technologies each present trade-offs. Monolithic sensors offer high spatial and time resolution with a minimal material budget, but have a limited radiation hardness and data rate capability~\cite{malta}. 3D sensors provide good radiation hardness and time resolution, but are limited by a reduced fill factor and an increased capacitance~\cite{cnm3d,timespot2020}. Low Gain Avalanche Detectors (LGADs) reach a high time resolution through internal gain~\cite{PELLEGRINI201412}, but their gain layer degrades under irradiation due to acceptor removal, which limits their lifetime in high-radiation environments~\cite{Mazza_2020}. Their granularity is also limited, as the termination structures between adjacent gain layers leave regions without gain. Inverse~\cite{PELLEGRINI201624}, trench-isolated~\cite{Paternoster2020TrenchIsolatedLG}, and resistive~\cite{8846722} LGADs aim to address this limitation, at the cost of additional process complexity. 
An ideal sensor would combine a fine pitch, a high fill factor, a radiation hardness comparable to that of thin planar sensors, and a good time resolution. A path towards these requirements would be a thin sensor in which a localised and radiation-hard gain region is implemented within each pixel. A small thickness limits the lateral spread of the collected charge and the drift time, while the multiplication compensates for the reduced deposited charge.
 
\subsection{The SiEM Principle}
The Silicon Electron Multiplier (SiEM) implements such a localised gain region without a doped gain layer. Instead, it uses a composite electrode structure embedded in the silicon bulk and fabricated with Micro-Electro-Mechanical Systems (MEMS) techniques. A potential difference applied across the electrodes creates a high electric field region. Charge carriers generated in the bulk by a Minimum Ionising Particle (MIP) drift into this region, where they undergo impact ionisation and create secondary electron-hole pairs. The multiplication is therefore defined by the geometry and the applied potentials rather than by a doping profile. The drift of the secondary charges induces the signal on the readout electrode.
 
\begin{figure*}[!t]
    \centering
    \captionsetup[subfigure]{labelformat=empty}
    \includegraphics[width=0.95\textwidth]{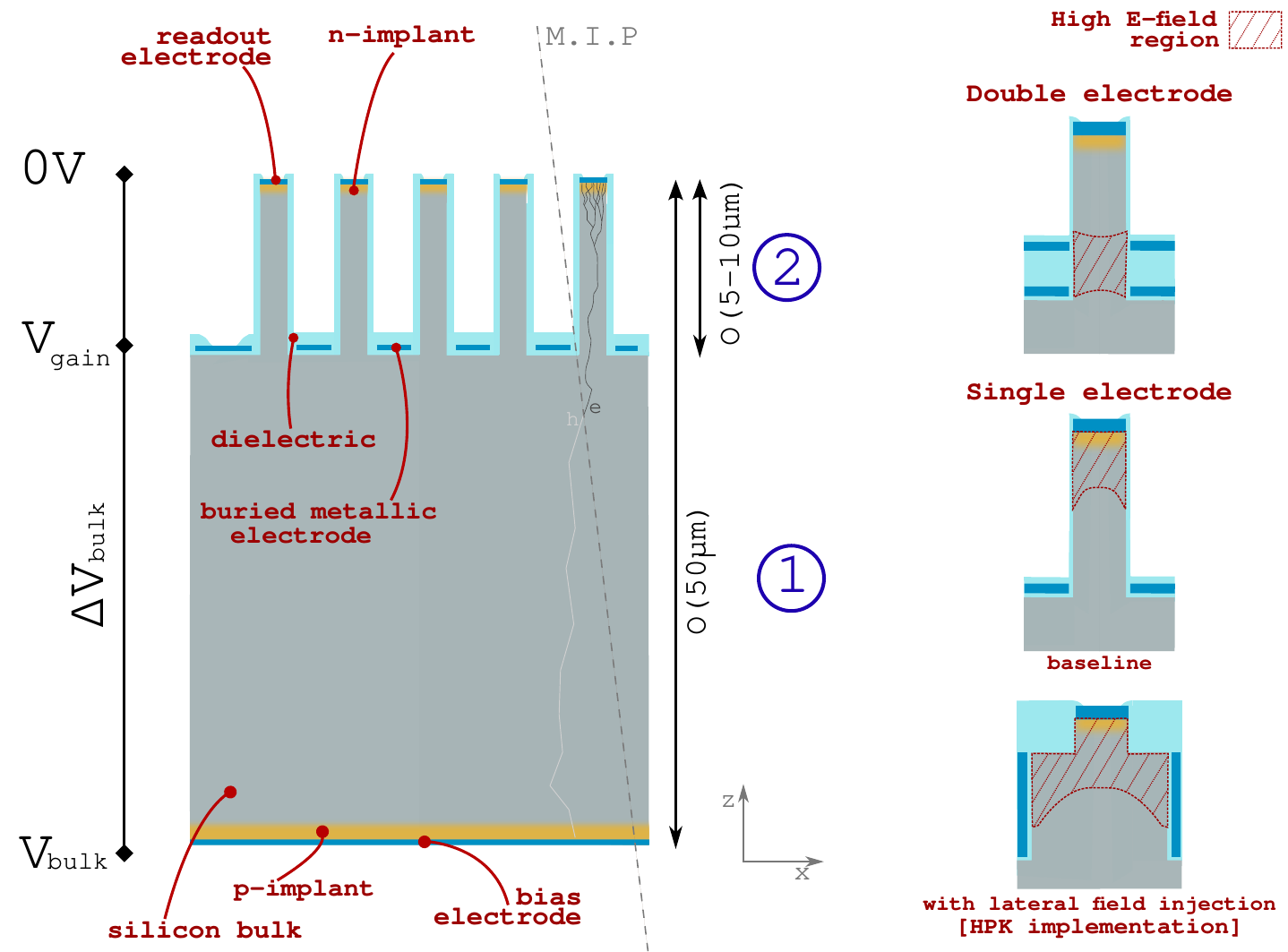}
    \caption{Left: schematic of a single electrode SiEM. It consists of an O(\SI{50}{\micro\meter}) thick depleted silicon region \drift adjacent to a region \induction made of silicon pillars, between which metallic electrodes are buried. The electrodes produce a high electric field region in \induction. Primary electrons produced by a traversing particle in \drift drift towards \induction, where they are multiplied and induce a signal while drifting towards the readout electrode. Right: types of buried electrodes, with at the top the double-electrode configuration, studied mainly in~\cite{HALVORSEN2022167325}, where the high electric field region is located between two buried electrodes, and below the single-electrode configuration, where the high electric field is located in the pillar close to the readout electrode. The middle picture shows the single-electrode configuration as discussed in~\cite{HALVORSEN2022167325,HALVORSEN2024169046,garfield-paper}, while the bottom one shows the evolution introduced by HPK in~\cite{hpk-paper}, with lateral field injection to both ease production and increase the field strength in the pillar, known as two-stage mesa-type SiEM.}
    \label{fig:siem-configurations}
\end{figure*}

This principle can be realised with different electrode geometries. Two main configurations have been studied~\cite{HALVORSEN2022167325}. The principle of the SiEM and the various electrode geometries are illustrated in Figure~\ref{fig:siem-configurations} and described below. 

In the double-electrode configuration, a high electric field is generated at the bottom of the pillar by applying a potential difference between the two buried electrodes. In the single-electrode configuration, the field is created between the amplification electrode and the readout electrode by applying a potential \vgain. The high field region is then located near the top of the pillar, close to the readout electrode. 

The two configurations also differ in their signal formation, which depends on where the multiplication occurs with respect to the readout electrode. In the double-electrode configuration, the multiplication takes place at the bottom of the pillar, away from the readout electrode, and the signal arises almost exclusively from the amplified electrons drifting through the pillar towards it. In the single-electrode configuration, the multiplication takes place close to the readout electrode, so that the amplified electrons are collected almost immediately and contribute only briefly, while the signal is dominated by the holes drifting back out of the pillar and across the bulk.

Owing to its simpler fabrication, the single-electrode configuration was selected for the first demonstrator productions.
 
\subsection{Initial Fabrication Approaches}
Following conceptual and simulation studies, several fabrication routes have been explored to produce SiEM demonstrators.
 
Studies have been carried out with the Centro Nacional de Microelectrónica (CNM), Barcelona, within the AIDA-innova programme. Several process steps have been developed, including the Deep Reactive Ion Etching (DRIE) parameters needed to reach the target geometry in a repeatable and scalable way, and a method to pattern the metal layer reliably despite the height differences across the structure. The production wafers have been doped, and the demonstrator structures are expected to be available soon~\cite{AIDAinnova}.
In parallel, an exploratory collaboration with the Paul Scherrer Institute (PSI) investigated Metal Assisted Chemical Etching (MACEtch) as an alternative fabrication method~\cite{metaletch}. This technique potentially offers lower cost and simpler processing, and can reach higher pillar aspect ratios, which suits the single-electrode configuration. Test structures were produced and showed that the diode behaviour is preserved after etching~\cite{HALVORSEN2024169046}. However, the leakage current was too high due to the formation of a Schottky diode at the interface between the platinum etching catalyst and the silicon. Future iterations will require an oxide layer between the etching metal and the amplification electrode. This approach therefore needs further development before its effectiveness can be assessed.
 
\subsection{Hamamatsu Photonics Demonstrator Production}
\label{sec:hpkprod}
In parallel with these efforts, Hamamatsu Photonics (HPK) produced functional SiEM demonstrators, building on the work described in~\cite{HALVORSEN2022167325}. HPK implemented a single-electrode configuration with a two-stage mesa-type geometry, which simplifies the production process while allowing a high electric field to be reached in the pillars at lower voltages~\cite{hpk-paper}.

\begin{figure}
    \centering
    \captionsetup[subfigure]{labelformat=empty}
    \subfloat[\centering]{\includegraphics[width=0.95\columnwidth]{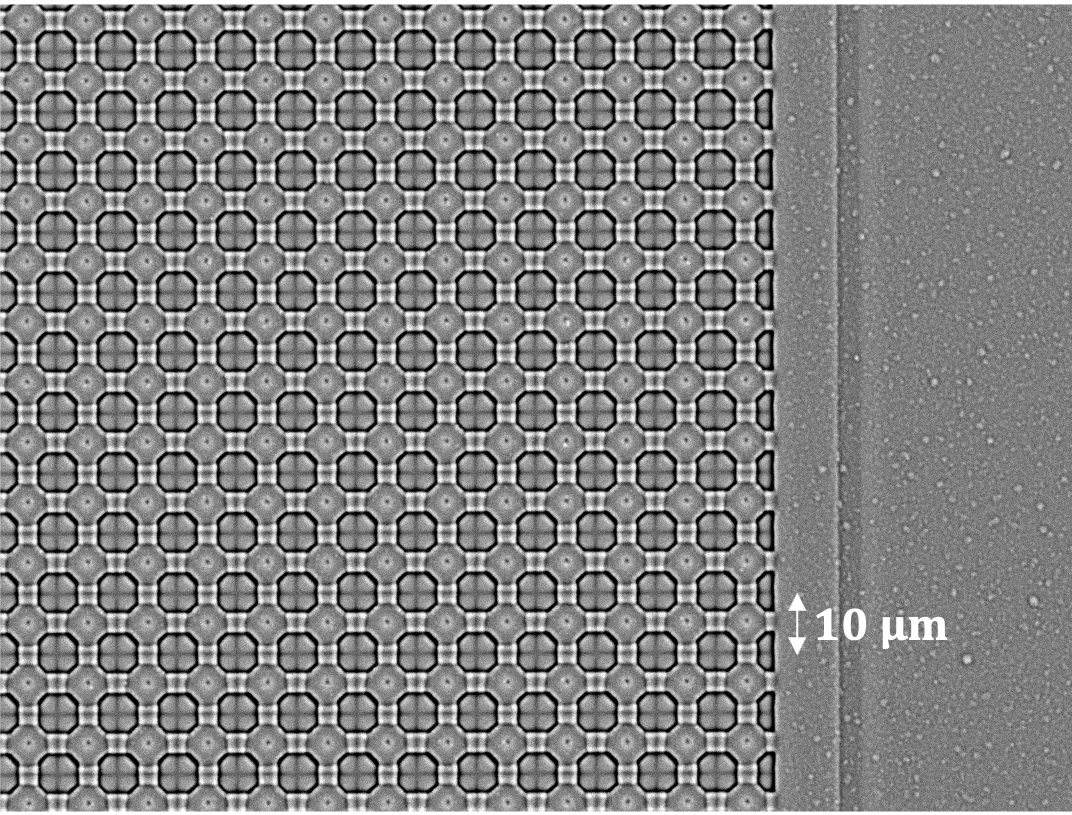}}\\[0.2em]
    \subfloat[\centering]{\includegraphics[trim = 0 0 0 0, clip,width=1\columnwidth]{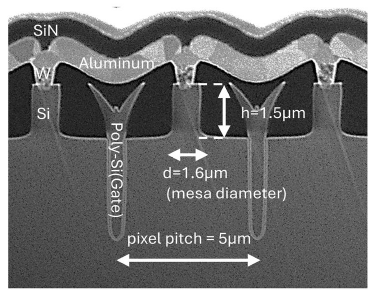}}
    \caption{Top: electron microscope image of the surface of an HPK sensor, showing the pixelated structure and readout electrode interconnecting all the pillars. Bottom: annotated image of the cross-section of the HPK geometry, showing the pillar height $h$, the pillar diameter $d$, the pitch $p$, the bulk thickness $B = \SI{50}{\micro\meter}$ (not to scale), the depth of the electric-field-injection region $D \approx \SI{4}{\micro\meter}$, and the electrodes at which the voltages are applied. The values of $h$, $d$ and $p$ for the two structures studied in this paper are given in the text. Reproduced from~\cite{hpk-paper}.}
    \label{fig:hpk-geometry}
\end{figure}

An example HPK demonstrator is shown in Figure~\ref{fig:hpk-geometry}. The gain electrode is a vertical polysilicon electrode, buried in oxide-filled trenches that surround each pixel. It forms an electric-field-injection region about \SI{4}{\micro\meter} in depth, whose lateral dimensions match the pixel pitch. This region is topped by a smaller cylindrical pillar of a few micrometres in height, the electron-multiplication region, with the readout electrode at its tip. As in the single-electrode configuration, the electric field is generated between the buried gain electrode and the readout electrode but here the injection region focuses the charge carriers laterally into the pillar, while the field is concentrated vertically in it, which produces the locally high field needed for impact ionisation~\cite{hpk-paper}. The multiplication region is located at the tip of the pillar, immediately below the readout electrode, so the signal formation is expected to be that of the single-electrode configuration. In~\cite{hpk-paper}, the injection region and the pillar are referred to as the injection stage and the cylindrical mesa, and the gain and readout electrodes as the gate and the cathode.
 
All sensors have an active area of \SI{2}{\milli\meter}~$\times$~\SI{2}{\milli\meter} and a bulk thickness of \SI{50}{\micro\meter}. All pixel readout electrodes of a sensor are interconnected, which allows a first characterisation without a pixelated readout. Two structures are studied in this paper. The \siemfive has 400~$\times$~400 pixels at a pitch of \SI{5}{\micro\meter}, with a pillar diameter $d = \SI{1.6}{\micro\meter}$ and a pillar height $h = \SI{1.5}{\micro\meter}$. It corresponds to the reference device described in~\cite{hpk-paper}. The \siemten has 200~$\times$~200 pixels at a pitch of \SI{10}{\micro\meter}, with $d = \SI{2}{\micro\meter}$ and $h = \SI{1.5}{\micro\meter}$. The \SI{10}{\micro\meter} pitch structure is not discussed specifically in~\cite{hpk-paper} but belongs to the same production.
 
Three voltages are used throughout this paper. The gain voltage \vgain is the potential difference between the gain electrode and the readout electrode, and sets the field in the pillar. The bulk voltage \deltavbulk is the potential difference between the backside and the gain electrode, and depletes the bulk. The total voltage applied between the backside and the readout electrode is therefore \vbulk = \deltavbulk + \vgain, as illustrated in Figure~\ref{fig:siem-configurations}. In the following, the voltages are quoted as absolute values, but the backside and the gain electrode are biased negatively with respect to the readout electrode.\\

This paper presents the characterisation of these demonstrators. Section~\ref{sec:simulation} uses Sentaurus TCAD and Garfield++ simulations of the two structures to describe their expected behaviour: the depletion of the bulk, the amplification of the leakage current in the pillar, the onset of gain and of the breakdown, and their dependence on the geometry, as well as the charge collection efficiency for MIPs. Section~\ref{sec:electrical} presents the I-V characteristics measured at various temperatures and bulk voltages. They are compared with the simulation, and the relative current increase with \vgain is used as a proxy for the gain to study its dependence on the geometry, the bulk voltage and the temperature. The I-V characteristics alone, however, cannot establish that this current increase results from charge multiplication. Section~\ref{sec:testbeam} therefore reports test beam measurements with MIPs, in which the collected charge is measured as a function of \vgain and compared with that of a planar sensor of the same thickness, with the relative current increase, and with the simulation. A first measurement of the time resolution is also reported. Section~\ref{sec:conclusions} summarises the results and outlines the next steps.

\section{Simulation of the HPK Structure}
\label{sec:simulation}

Detailed TCAD simulations of the HPK geometry are presented in~\cite{hpk-paper}. The same model is used here for both the \siemfive and the \siemten, to describe the behaviour of the device at various operating points and to predict the features that are looked for in the measurements. The simulation follows the methodology described in~\cite{garfield-paper}. The Sentaurus TCAD device simulator~\cite{sentaurus}, version S-2023.12, is used with the PARDISO solver~\cite{pardiso-7.2d,pardiso-7.2e}, the Canali mobility model~\cite{canali}, the Slotboom band gap model~\cite{SLOTBOOM_1,SLOTBOOM_2,SLOTBOOM_3} and the van Overstraeten-de Man impact ionisation model~\cite{vanoverstreaten}. The 3D mesh is generated by the Sentaurus Structure Editor, with cell sizes ranging from 200 to \SI{10}{\nano\meter}.

At each operating point, the electric field, the weighting potential, and the impact ionisation rates are extracted from TCAD and imported into Garfield++~\cite{garfield} for a Monte Carlo simulation of the response to MIPs. The charge carriers are tracked as they drift and diffuse through the device, and the current induced on the readout electrode is computed with the Ramo-Shockley theorem~\cite{ramo,shockley}. Only a single pillar unit cell is considered, with periodic boundary conditions. The collected charge is obtained by integrating the induced current over time.

\subsection{I-V characteristics and onset of impact ionisation}
\begin{figure}
    \centering
\includegraphics[width=0.95\columnwidth]{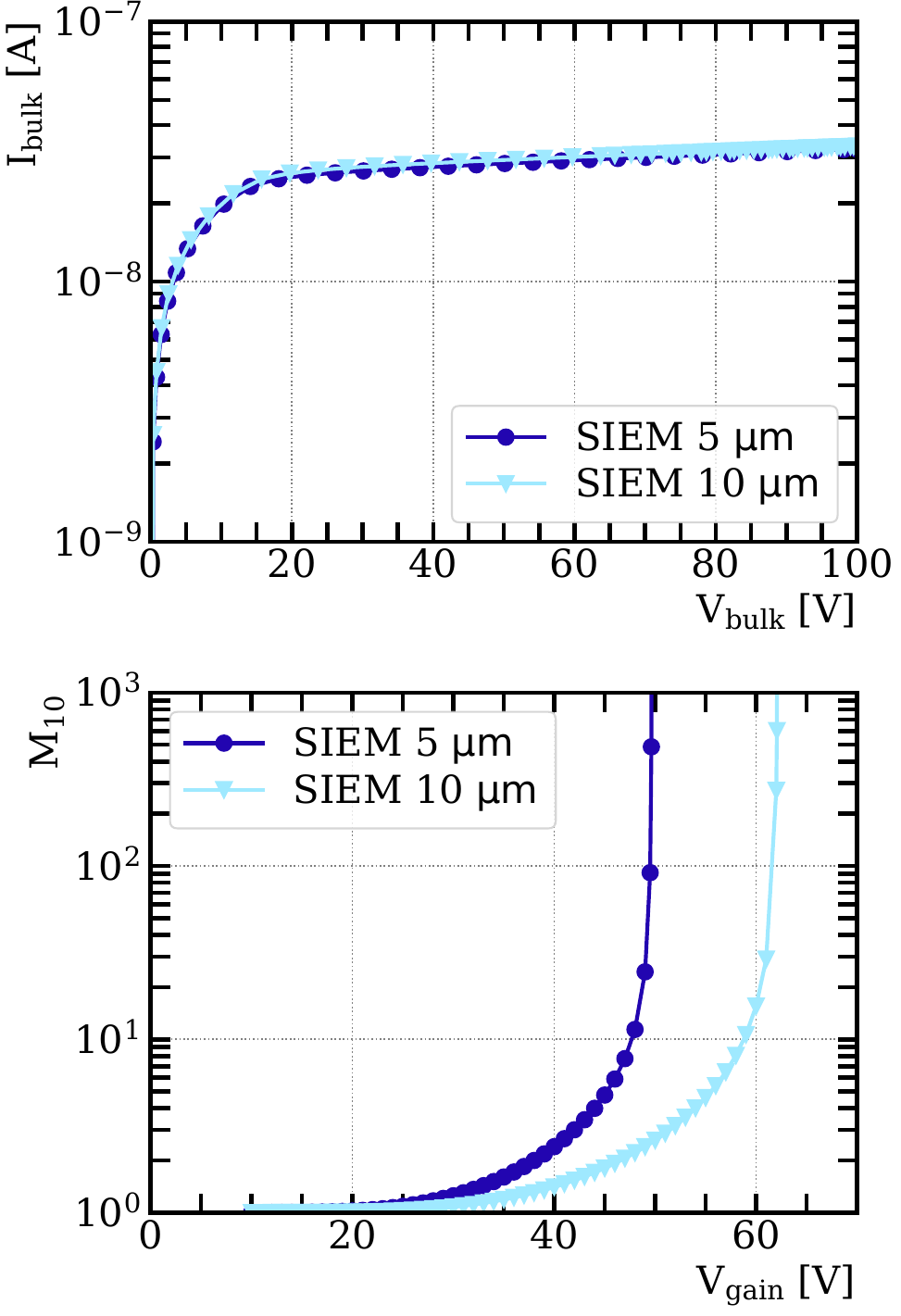}
    \caption{Simulated I-V characteristics for the two-stage biasing. Top: current as a function of \deltavbulk at \vgain $= \SI{0}{\volt}$, showing the diode behaviour. Bottom: current as a function of \vgain at \deltavbulk $= \SI{100}{\volt}$, showing increasing current from \SIrange{20}{30}{\volt} and the breakdown near \SI{45}{\volt} for the \siemfive, and increasing current from \SIrange{30}{40}{\volt} and the breakdown near \SI{60}{\volt} for the \siemten.}
    \label{fig:simulated-iv}
\end{figure}

The simulated I-V characteristics give the first indication of charge multiplication in the structure. The simulation follows a two-stage biasing protocol. In the first stage, \vbulk is ramped with \vgain at \SI{0}{\volt}, and the bulk depletes. The structure then behaves as a reverse-biased diode and follows the Shockley equation. In the second stage, \vgain is increased while \deltavbulk is kept at \SI{100}{\volt} to preserve full depletion. The current starts to rise from \vgain=\SIrange{20}{30}{\volt} for the \siemfive and from \vgain=\SIrange{30}{40}{\volt} for the \siemten, and reaches breakdown at about \SI{45}{\volt} and \SI{60}{\volt} respectively. The two regimes are shown in Figure~\ref{fig:simulated-iv}, as a function of the total voltage \vbulk applied across the device.

This current increase is the multiplication of the leakage current in the pillar. The carriers thermally generated in the depleted bulk drift towards the pillar along the same field lines as the electrons produced by a MIP, and are multiplied in the same high field region. The relative increase of the current with \vgain therefore measures the multiplication that the structure applies to any charge reaching the pillar, and can be used as a proxy for the gain. This proxy is defined in the following as the ratio of the current at a given \vgain to the current at $V_{gain} = \SI{10}{\volt}$, where the impact ionisation in the pillar is still negligible, once the bulk is fully depleted:

\begin{equation}
    M_{10} = \frac{I(V_{\mathrm{gain}})}{I(V_{\mathrm{gain}} = \SI{10}{\volt})}.
    \label{eq:m10}
\end{equation}

A similar quantity is used in~\cite{hpk-paper}, where the amplified current is the photocurrent produced by a \SI{905}{\nano\meter} continuous-wave laser focused to a spot of \SIrange{10}{20}{\micro\meter} at the top of the sensor, rather than the leakage current, and where the normalisation is taken at \vgain $= \SI{0}{\volt}$ at a fixed backside voltage. The two definitions are therefore similar in nature but not identical, and their values are not directly comparable. The normalisation at \SI{10}{\volt} is also convenient for the measurements, as discussed in Section~\ref{sec:electrical}.

The proxy defined in Eq.~\eqref{eq:m10} has two limitations. Firstly, it only measures multiplication as long as the current is dominated by the thermally generated carriers that reach the pillar. As the breakdown develops, the current is no longer a measure of that multiplication. Secondly, it assumes that no other current contribution increases with \vgain. This assumption can be verified in the simulation, where every contribution is known, but not in a measured I-V characteristic, where a current induced by a process defect or by a surface effect would be indistinguishable from the multiplied leakage current. The response of the structure to the charge deposited by a MIP is therefore studied as well, in the simulation in Section~\ref{sec:simulation-cce} and in the test beam measurements of Section~\ref{sec:testbeam}, in order to establish the multiplication of a charge of known origin and deposited at a known position.

\subsection{Electric field in the pillar}
\begin{figure}
    \centering
    \includegraphics[width=0.99\columnwidth]{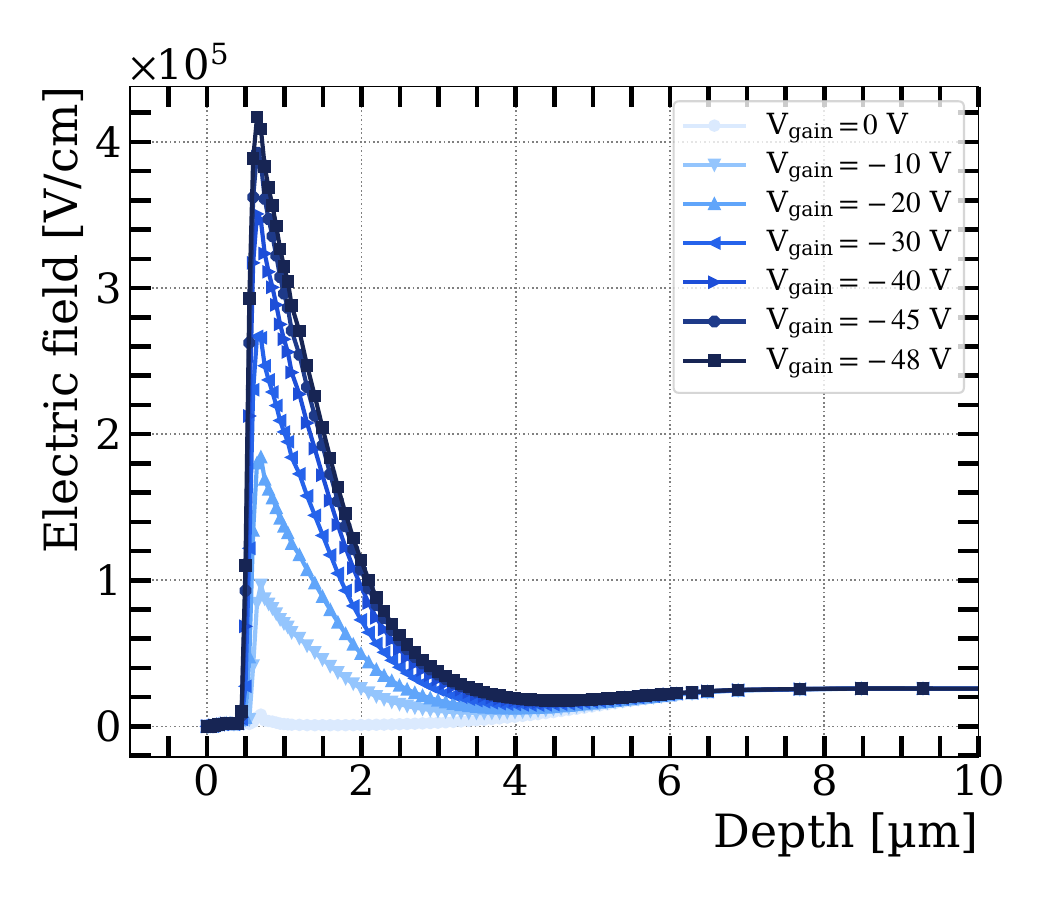}\\[0.2em]
    \includegraphics[width=0.98\columnwidth]{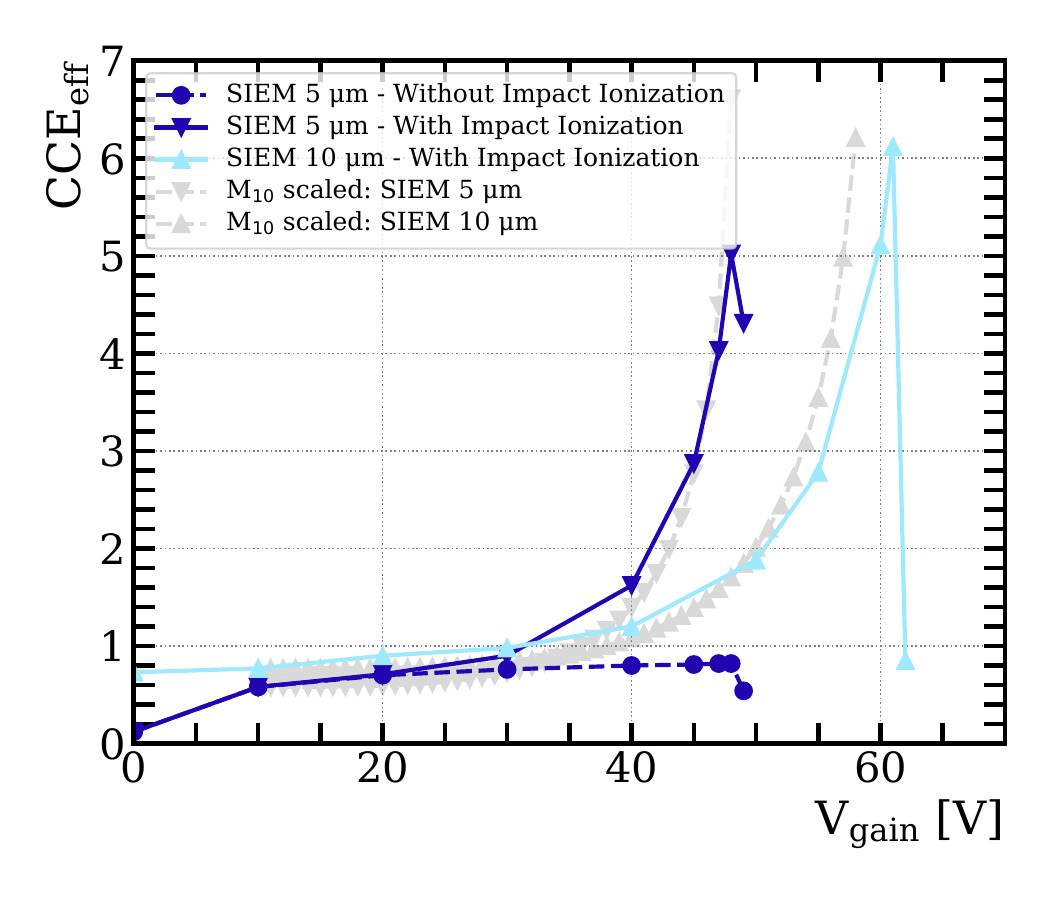}
    \caption{Top: electric field magnitude along the pillar depth ($x$-axis) at the pillar centre, for \vgain $= \SIlist{10;20;30;40}{\volt}$ for the \siemfive. The readout electrode is at \SI{0}{\micro\m}. Bottom: CCE as a function of \vgain for tracks through the pillar centre. The \siemfive and the \siemten are shown with the impact ionisation model enabled, and the \siemfive also with the impact ionisation model disabled, in which case the CCE saturates near unity when the transport is maximally efficient.}
    \label{fig:electric-field}
\end{figure}

The distribution of the electric field in the pillar shows how the device behaviour evolves with \vgain. Figure~\ref{fig:electric-field} (top) shows the field magnitude along the pillar depth, at its centre, for several values of \vgain for the \siemfive. At low gain voltage (\vgain $< \SI{20}{\volt}$) the field remains weak along the whole pillar. The carriers therefore drift slowly and can be trapped before reaching the readout electrode, and the charge is only partly collected as shown in Figure~\ref{fig:electric-field} (bottom). As \vgain increases beyond \SIrange{20}{30}{\volt}, the field strengthens over the full pillar depth and the high-field amplification region develops near the readout electrode. A \vgain above this threshold is therefore needed not only to initiate impact ionisation, but also to transport the charge efficiently through the pillar.

\subsection{Charge collection efficiency}
\label{sec:simulation-cce}
The Charge Collection Efficiency (CCE) is studied with Garfield++ simulations of MIPs crossing the sensor. It is defined as the ratio of the charge collected at the readout electrode to the charge deposited by the MIP along its trajectory through the \SI{50}{\micro\meter} active thickness~\cite{garfield-paper}. With this definition, the CCE contains both the probability that the carriers reach the readout electrode, or the transport efficiency $\varepsilon$, and the gain $G$ from the multiplication in the regions where impact ionisation occurs, so that CCE~$= \varepsilon \times G$.

Figure~\ref{fig:electric-field} (bottom) shows the CCE as a function of \vgain, for tracks through the pillar centre. Three regimes are visible. The CCE first rises with \vgain between \SI{0}{\volt} and \SI{10}{\volt}, as the field in the pillar becomes strong enough to transport the charge to the readout electrode. It then reaches a plateau, where the transport is efficient but the field is still too low for impact ionisation. It finally rises again when the field in the pillar becomes high enough to multiply the charge. The onset of this last rise occurs at a higher \vgain for the \siemten than for the \siemfive. When the impact ionisation model is disabled, the \siemfive curve saturates near unity instead, which confirms that the third regime occurs due to multiplication and not from an improvement of the transport.

\begin{figure}
    \centering
    \includegraphics[width=0.99\columnwidth]{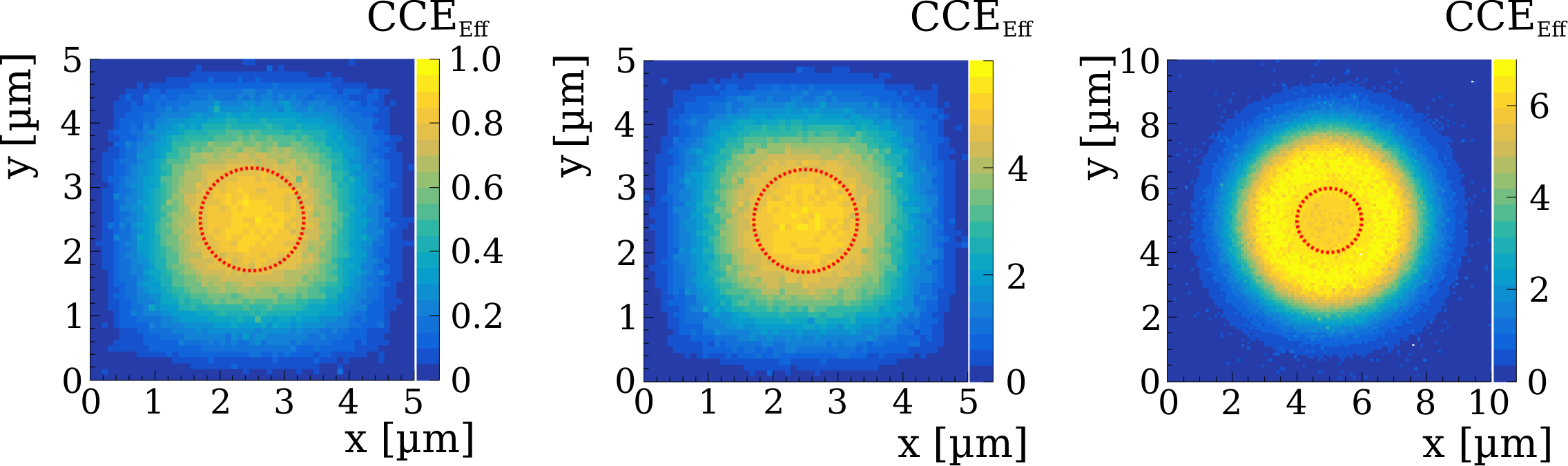}
    \includegraphics[width=0.99\columnwidth]{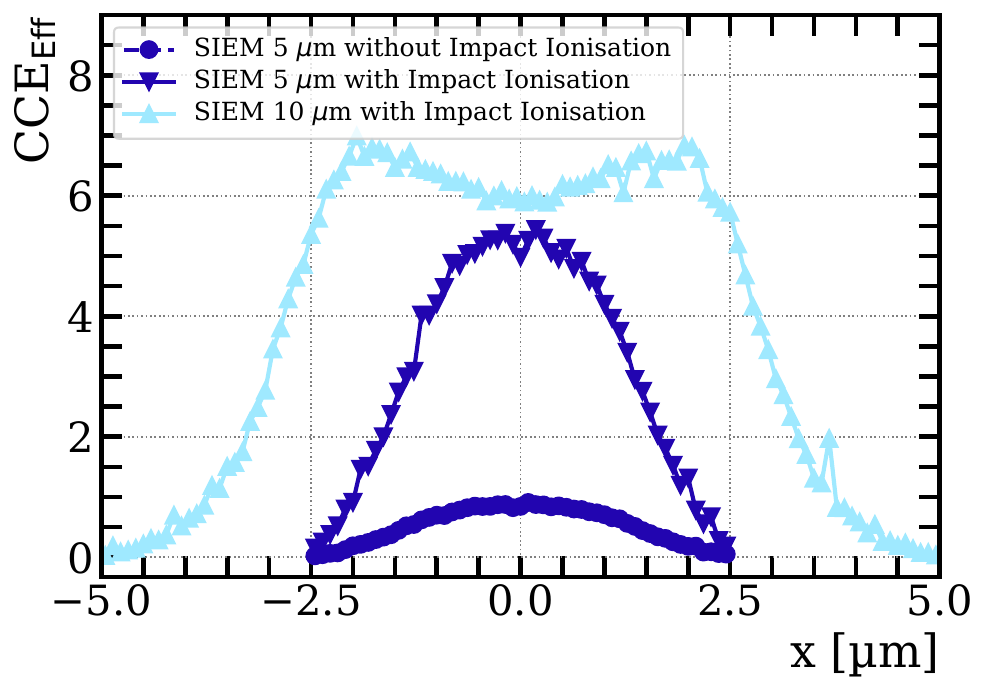}
    \caption{CCE as a function of the MIP entry position in the pixel, at the maximum \vgain before breakdown (\SI{48}{\volt} for the \siemfive and \SI{61}{\volt} for the \siemten). Top panel: two-dimensional maps in the $x$-$y$ plane; left map: \siemfive with the impact ionisation disabled; middle map: \siemfive with the impact ionisation enabled; right map: \siemten with the impact ionisation enabled. The pillar position is indicated by the red circle in each map. Bottom panel:  one-dimensional cross-section along the $x$-axis at the pixel centre, for the three configurations of the top panel. }
    \label{fig:cce-gain}
\end{figure}

The transport component is further studied with the impact ionisation disabled for the \siemfive case. Two behaviours then appear depending on the trajectory of the MIP. For particles going through a pillar, the transport efficiency is driven by the field strength inside the pillar. For particles passing outside the pillar, it also depends on the lateral drift towards the pillar, that is on the focusing of the charge into the amplification region. This is illustrated in Figure~\ref{fig:cce-gain}, which shows the CCE as a function of the MIP entry position in the pixel at the maximum \vgain before breakdown of \SI{48}{\volt} for the \siemfive and \SI{61}{\volt} for the \siemten in these simulations.
With the impact ionisation disabled (left map of the top panel), the collection is reduced away from the pillar, where the lateral drift distances are large and the focusing is less effective, and the one-dimensional cross-section along $x$ saturates below unity even at this voltage. With the impact ionisation enabled, the $\varepsilon \times G$ product exceeds unity in the pillar region for both structures and reaches about 5 for the \siemfive (middle map of the top panel) and about 6 for the \siemten (right map of the top panel) at the pillar centre, where the transport efficiency is close to unity. The response remains non-uniform across the pixel, since the charge collected far from the pillar is still incomplete. The structures studied in the measurements have all their pixels interconnected, such that the measured charge averages over the whole pixel area and is expected to be lower than these pillar-centre values.

To summarise, these simulation results indicate which features should be looked for in the measurements. The current should follow a diode behaviour while the bulk depletes, then rise with \vgain from about \SIrange{20}{30}{\volt} for the \siemfive and from about \SIrange{30}{40}{\volt} for the \siemten, reaching breakdown at about \SI{45}{\volt} and \SI{60}{\volt} respectively. The relative current increase $M_{10}$ defined in Eq.~\eqref{eq:m10} should reproduce this behaviour. The charge collected from MIPs is expected to remain small at \vgain $= \SI{10}{\volt}$, where the transport through the pillar is incomplete, and then exceed the charge deposited in \SI{50}{\micro\meter} of silicon as \vgain approaches breakdown, by a factor of up to about 5 and 6 at the pillar centre for the \siemfive and the \siemten respectively.

\section{Electrical Characterisation}
\label{sec:electrical}
 
The electrical characterisation of the HPK demonstrators follows the two-stage biasing protocol used in the simulation. The bulk voltage \vbulk is first raised to \SI{100}{\volt} to deplete the bulk, while \vgain is kept at \SI{0}{\volt}. \vgain is then swept until the current reaches \SI{1}{\micro\ampere}, with \deltavbulk kept constant at \SI{100}{\volt}. The current is recorded at each voltage step to build the I-V characteristic, from which the relative current increase $M_{10}$ defined in Eq.~\eqref{eq:m10} is obtained.
 
\subsection{Experimental setup}
The measurements are performed with a custom setup. Three remotely movable probe needles are used to bias and read out the samples and are positioned on the sensor pads. Two Keithley 2410 sourcemeters with a shared ground provide the voltages.
 
The sample is held on a metal block mounted on a Peltier element, cooled by a recirculating chiller, allowing to reach temperatures down to \SI{-30}{{\celsius}}. The setup is placed in a sealed chamber in which the temperature and the humidity are monitored to avoid condensation on the sensor at low temperatures. The reproducibility was verified by repeating the measurement on the same sample after the maximal current had been reached, which is beyond the breakdown voltage. No damage was observed from this procedure as the results were identical.
 
\subsection{Yield and homogeneity}
About 20 samples of each structure type were measured at room temperature. Around 80\% of them show the expected diode and multiplication behaviour. On the remaining structures, scratches are visible on the sensor surface, which indicates the behaviour is due to handling damage rather than to a production issue.
 
The I-V characteristics are shown in Figure~\ref{fig:iv-all-types} (top). The behaviour during the bulk ramp is similar for all structures, with a current below \SI{1}{\nano\ampere} at depletion. When \vgain is then raised from \SIrange{0}{10}{\volt}, the current increases to about \SI{10}{\nano\ampere} for the \siemfive structures, and slightly less for the \siemten. Normalising at \vgain $= \SI{10}{\volt}$, as in Eq.~\eqref{eq:m10}, removes this contribution from the comparison between structures, as shown in Figure~\ref{fig:iv-all-types} (bottom).

Within each structure type, the behaviour is homogeneous. The breakdown voltage $V_{\mathrm{gain}}^{\mathrm{BD}}$ spreads by at most \SI{5}{\percent} for the \siemfive and \SI{6}{\percent} for the \siemten, which can limit the maximum gain accessible on a given structure. 
 
\subsection{Comparison with the simulation}
The measured I-V characteristics are compared with the TCAD simulation in Figure~\ref{fig:iv-all-types}. During the bulk ramp, at \vgain $= \SI{0}{\volt}$, the measured and simulated currents agree in shape and in the voltage at which full depletion is reached. The two differ in one feature only: the step observed when \vgain is raised from \SIrange{0}{10}{\volt}, which shifts the measured current with respect to the simulated one, and which the simulation does not reproduce. The TCAD model does not include a calibrated model of the generation current at the Si/SiO$_2$ interfaces~\cite{hpk-paper}, and this contribution is therefore not expected to be described. A rapid rise of the dark current at low gain voltage is also reported in~\cite{hpk-paper}, where it is attributed to a gate-induced modulation of the leakage current at the interface under the gate oxide, and not to multiplication in the high-field region. Normalising at \vgain $= \SI{10}{\volt}$, as in Eq.~\eqref{eq:m10}, removes this contribution from the comparison between structures.

Above this step, the measurements follow the simulation. The simulated $M_{10}$ curves reproduce the shape of the measured ones over the whole range, from the onset of the current increase to the breakdown, and are in good agreement with them: the difference in \vgain between the simulation and the average of the measured samples, taken at the same value of $M_{10}$, is of the order of \SIrange{6}{12}{\percent}. The simulation therefore describes the depletion, the onset of the multiplication and the breakdown, which are the quantities used in the following.
 
\subsection{Dependence on the geometry}
\begin{figure}
    \centering
    \includegraphics[width=1\columnwidth]{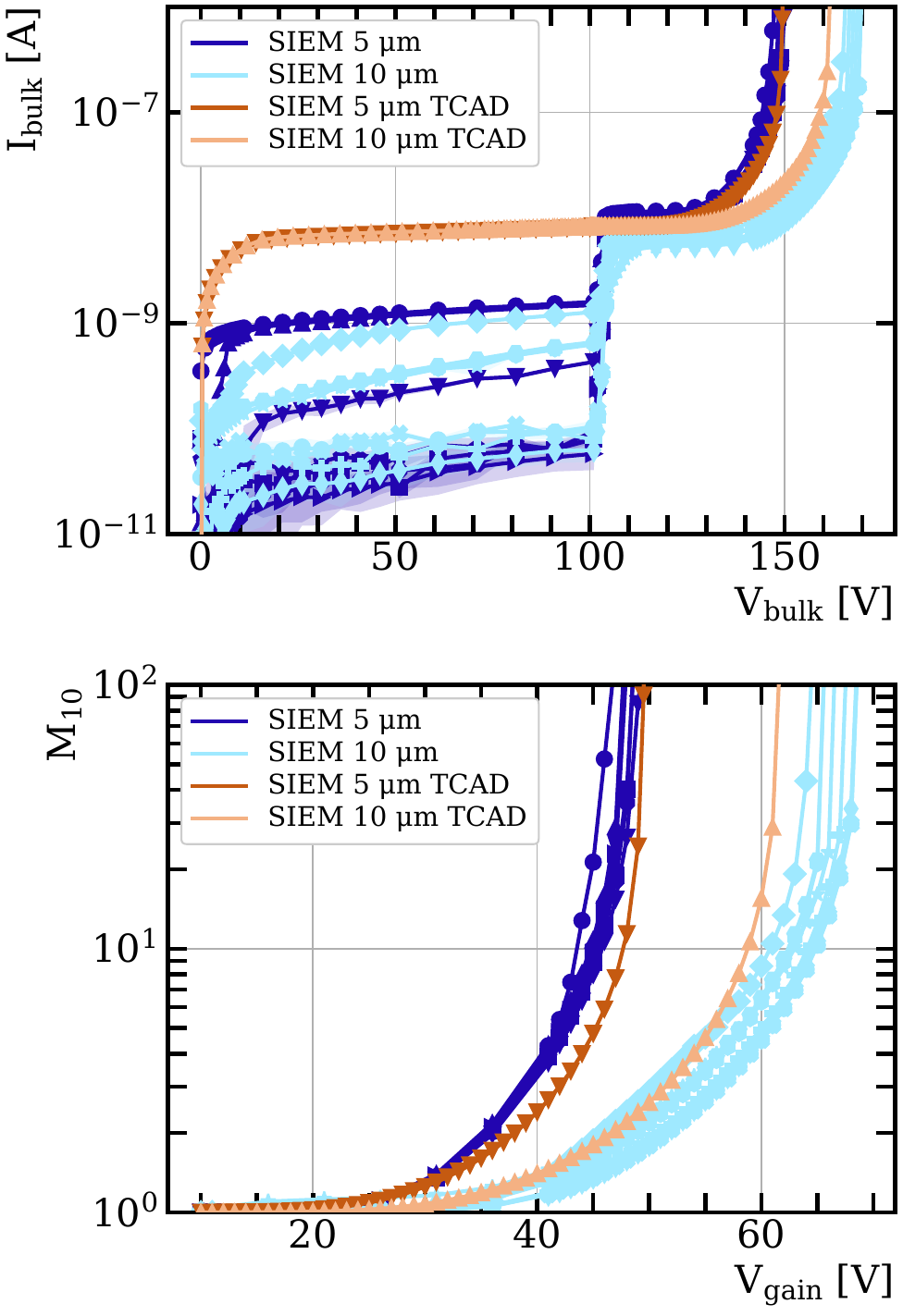}
    \caption{Top: current as a function of \vbulk for the \siemfive and \siemten structures at room temperature, showing the similar behaviour during the bulk ramp and the current step when \vgain is raised from \SIrange{0}{10}{\volt}. The TCAD simulation of the \siemfive is shown for comparison. Bottom: relative current increase $M_{10}$, defined in Eq.~\eqref{eq:m10}, as a function of \vgain for the same structures. Different markers distinguish the individual samples of a given type. The onset of the current increase occurs at a higher \vgain for the \siemten than for the \siemfive.}
    \label{fig:iv-all-types}
\end{figure}

The evolution of $M_{10}$ with \vgain is shown in Figure~\ref{fig:iv-all-types} (bottom). 

A given value of $M_{10}$ is reached at a lower \vgain for the \siemfive than for the
\siemten, as expected from the simulation. In~\cite{hpk-paper}, the dependence on the
pillar diameter $d$ is studied at a fixed pitch of \SI{5}{\micro\meter}: a smaller $d$
concentrates the field lines into a smaller area, which increases the field in the pillar
and therefore the impact ionisation. Here, the pitch differs between the two structures
and two effects come into play. The ratio between the area of the gain electrode, which
defines the field injection region, and the area of the pillar is larger for the \siemten.
Taken alone, this would increase the field in the pillar and reduce the potential needed
to reach a given $M_{10}$. The distance between the gain electrode and the readout
electrode is also larger, which increases the potential needed to reach a given field. The
two effects cannot be separated with a single geometry per pitch, but the second effect appears to dominate as the \siemten requires a larger potential to reach the same $M_{10}$.

\subsection{Dependence on the bulk voltage and on the temperature}
\begin{figure}
    \centering
    \includegraphics[width=1\columnwidth]{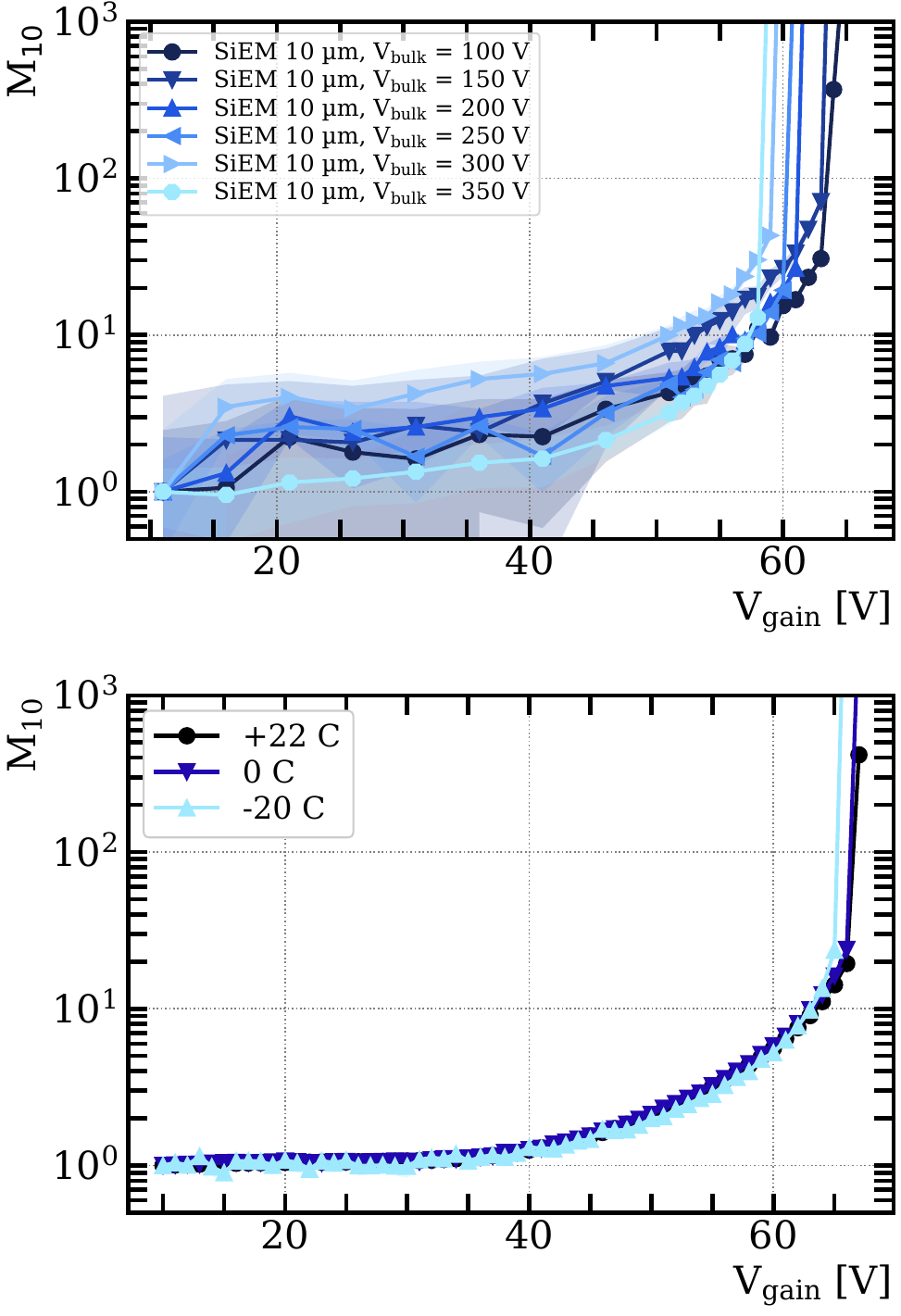}
    \caption{Relative current increase $M_{10}$ as a function of \vgain for \siemten samples. Top: for \deltavbulk from \SIrange{100}{350}{\volt} in steps of \SI{50}{\volt}, at \SI{-30}{\celsius}. Bottom: for temperatures of \SI{-20}{\celsius}, \SI{0}{\celsius} and \SI{22}{\celsius}, at \deltavbulk $= \SI{100}{\volt}$. The curves overlap in both cases, except for the breakdown point.}
    \label{fig:iv-temperature}
\end{figure}

The dependence on the bulk voltage was measured on \siemten samples at \SI{-30}{\celsius}, by stepping \deltavbulk from \SIrange{100}{350}{\volt} in steps of \SI{50}{\volt} and repeating the \vgain scan at each step. The resulting $M_{10}$ curves are shown in Figure~\ref{fig:iv-temperature} (top). They overlap, except for the breakdown point, which moves to lower \vgain as \deltavbulk increases. By extracting $V_{\mathrm{gain}}^{\mathrm{BD}}$ in the current range between \SI{e-8}{\ampere} and \SI{e-7}{\ampere}, the shift is shown to be approximately linear and amounts to about \SI{1}{\volt} per \SI{50}{\volt} of \deltavbulk.

 The dependence on the temperature was measured on \siemten samples at \SI{-20}{\celsius}, \SI{0}{\celsius}, and \SI{22}{\celsius}. The leakage current at depletion decreases from \SI{4.8}{\nano\ampere} at \SI{22}{\celsius} to \SI{0.11}{\nano\ampere} at \SI{-20}{\celsius}, that is by about a factor \num{1.09} per \SI{}{\celsius}, as expected for a generation current in the depleted silicon bulk and in agreement with the factor of about \num{1.1} per \SI{}{\celsius} reported in~\cite{hpk-paper}. The $M_{10}$ curves, shown in Figure~\ref{fig:iv-temperature} (bottom), overlap over the whole temperature range up to the vicinity of the breakdown. The absolute current therefore varies by more than a factor \num{40} while its relative increase with \vgain does not vary significantly. 
 
 The breakdown occurs at a lower \vgain as the temperature decreases, by about \SI{50}{\milli\volt\per\celsius} when extracted at $M_{10} = 20$, where the current increase is already dominated by the onset of the breakdown. This positive temperature coefficient is the expected signature of an avalanche breakdown~\cite{Ohsugi1994MicrodischargesOA}: as the temperature rises, phonon scattering shortens the carrier mean free path, so a higher field is required to reach the ionisation threshold. A breakdown dominated by band-to-band tunnelling would show the opposite behaviour, since the narrowing of the band gap with temperature makes tunnelling easier. The multiplication responsible for the rise of $M_{10}$ is therefore identified as impact ionisation.

Below the breakdown region, the ionisation coefficients are expected to retain the same weak temperature dependence, which would appear as a small displacement of the $M_{10}$ curves along \vgain, of a few tens of \si{\milli\volt\per\celsius} as reported in~\cite{hpk-paper}. Such a displacement could not be resolved with the present measurements. It is comparable to the precision with which $M_{10}$ is determined, and its extraction is sensitive to the normalisation at \SI{10}{\volt}, where the current is small at \SI{-20}{\celsius} and its relative uncertainty correspondingly large; depending on the value of $M_{10}$ at which it is extracted, the displacement is found with either sign. A measurement with a normalisation point better matched to the temperature range would be required to reproduce the coefficient of~\cite{hpk-paper}.
 
These measurements establish that the current increase with \vgain behaves as a multiplication of the leakage current, that it follows the simulation, and that it is reproducible across samples and stable against the bulk voltage and the temperature. They do not establish, however, that a localised charge is multiplied in the same way. As discussed in Section~\ref{sec:simulation}, $M_{10}$ is no longer a measure of the multiplication once the breakdown develops, and a \vgain-dependent current of another origin, such as a surface or defect-related contribution, cannot be separated from the multiplied leakage current in an I-V characteristic. The response to the charge deposited by MIPs is therefore measured in Section~\ref{sec:testbeam}.

\section{Test Beam Characterisation}
\label{sec:testbeam}
 
The measurements of Section~\ref{sec:electrical} show that the current increases with \vgain as expected from the multiplication of the leakage current, but they cannot establish that a localised charge is multiplied in the same way. The response of the structures to the charge deposited by MIPs is therefore measured, by exposing them to a particle beam and recording their transient response.
 
\subsection{Experimental setup}
The measurements were performed at the CERN SPS H8 beam line, with a mixed hadron beam of \SI{180}{\giga\eV/c}. The setup consists of two sensor planes placed one after the other along the beam, about \SI{5}{\centi\meter} apart. Each plane holds a sensor mounted on an OPTIMA board~\cite{optima-ref}, which serves as the preamplifier, inside an enclosure in which the temperature and the humidity are kept constant and that can be positioned with respect to the beam with motorised stages.
 
The first plane holds a Low Gain Avalanche Detector (LGAD) matrix of \num{5}~$\times$~\num{5} pixels~\cite{CMSLGAD}. Nine of these \SI{1.3}{\milli\meter}~$\times$~\SI{1.3}{\milli\meter} pixels, arranged in a \num{3}~$\times$~\num{3} matrix, were read out by an OPTIMA. The central pixel provided the trigger, and therefore defines a Region Of Interest (ROI) through which a charged particle has passed, while its neighbours were used for the fine alignment of the two planes with respect to each other. The second plane holds the Device Under Test (DUT), which was either a SiEM structure, of \SI{2}{\milli\meter}~$\times$~\SI{2}{\milli\meter} active area, or a circular planar diode of \SI{1}{\milli\meter} diameter from the production described in~\cite{Haimberger:2023fkr}, used to calibrate the OPTIMA channels.
 
The OPTIMA is a multichannel preamplifier system designed for the study of silicon sensor test structures, described in~\cite{optima-ref}. It provides a gain of about \SI{100}{\decibel} through two Transimpedance Amplifier (TIA) stages, with a low noise and a bandwidth suited to the fast signals of thin silicon sensors. Both OPTIMA outputs are connected to a Tektronix MSO64B oscilloscope operating at \SI{25}{\giga\sample\per\second}, which also handles the trigger logic based on the LGAD signal.
 
The alignment between the DUT and the ROI was checked for each configuration. For the SiEM, whose active area is larger than the ROI, the neighbouring pixels were used to verify that coincidences were seen with the SiEM all around the central pixel, which ensured that the ROI was fully contained in the SiEM active area. For the planar diodes, whose area is smaller than the ROI, the neighbouring pixels were used to verify that almost no coincidence was seen, which ensured that the diode was centred on the ROI. The response of the DUT recorded on an ROI trigger is therefore unbiased.

In the most recent datasets, when a new structure was installed, the OPTIMA was temporarily read out by a SAMPIC digitiser~\cite{sampic}, which allows the waveforms to be correlated with the tracks reconstructed by the Timepix4 telescope~\cite{tpx4_telescope}, confirming the quality of the alignment before taking the data with the oscilloscope.
 
\subsection{Charge calibration with planar sensors}
The OPTIMA amplifies the current through a TIA, so that the collected charge is proportional to the integral of the voltage waveform recorded by the oscilloscope. The proportionality constant was obtained from planar diodes of known thickness, \SIlist{100;200;300}{\micro\meter}, placed as the DUT plane and exposed to the same beam under the same conditions.
 
\begin{figure*}
    \centering
    \includegraphics[width=1\textwidth]{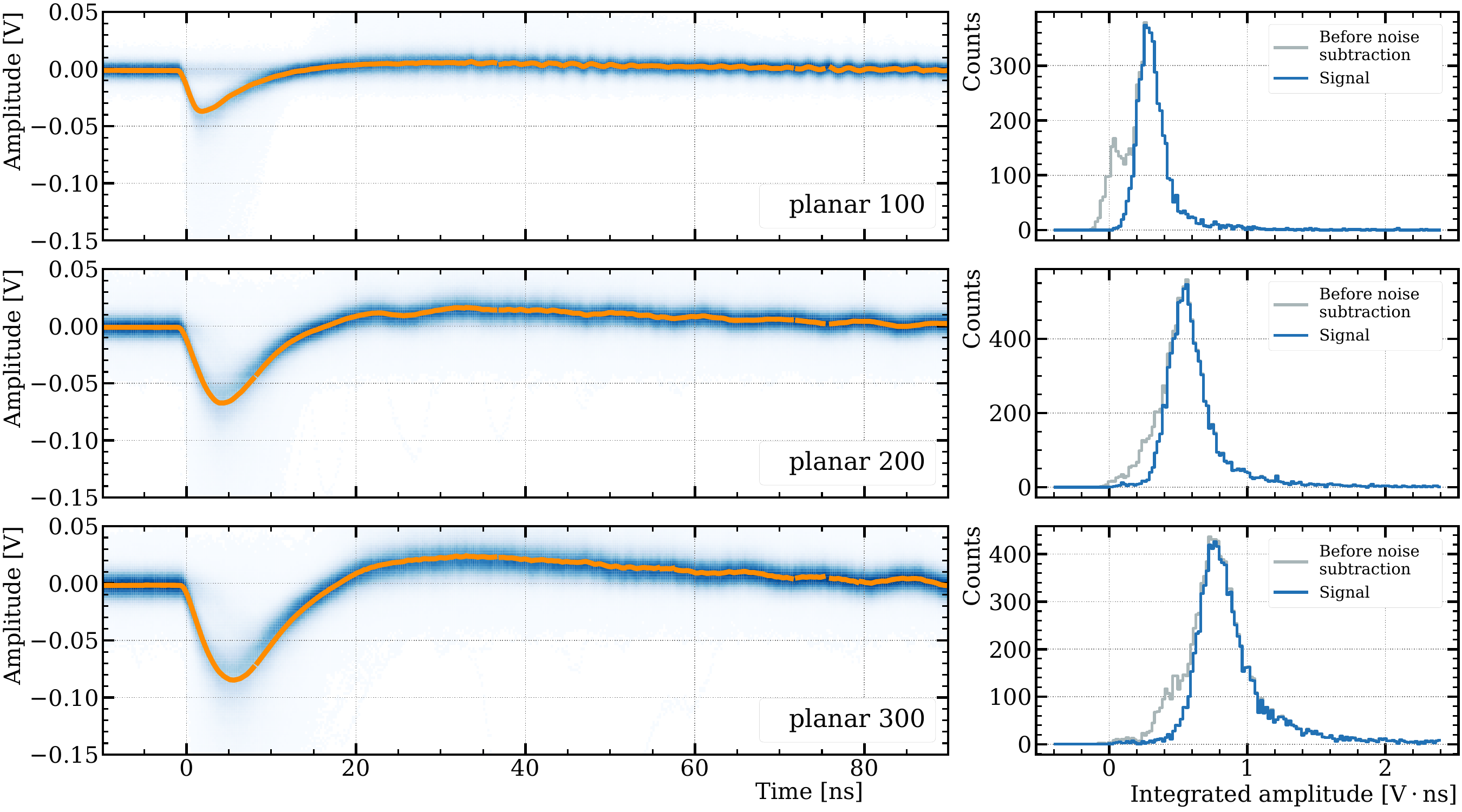}
    \caption{Left: waveforms recorded when the ROI triggers for the \SI{100}{\micro\meter}, \SI{200}{\micro\meter} and \SI{300}{\micro\meter} planar sensors, from top to bottom. The colour scale indicates the number of events in each time-voltage bin and the orange curve is the average waveform. Right: corresponding distributions of the waveform integral, in the same order, before and after the selection removing the events for which the particle missed the implant.}
    \label{fig:planar_waveforms}
\end{figure*}

For each sensor, the waveform is integrated from the start of the recording to the point at which the average signal returns to its baseline. Figure~\ref{fig:planar_waveforms} (left) shows the waveforms recorded when the ROI triggers for the three thicknesses, and Figure~\ref{fig:planar_waveforms} (right) shows the corresponding distributions of the waveform integral. Since the diodes are circular and smaller than the ROI, particles passing at the edge of the ROI may miss the implant. They then leave no signal, or only part of the charge is collected, as the weighting field extends beyond the implant footprint, in particular for the thicker sensors. These events are removed by requiring the waveform to be at least \SI{-20}{\milli\volt} at \SI{2}{\nano\second}, and the distributions before and after this selection are both shown in the figure.
 
The residual baseline contribution is estimated from pure noise events and subtracted from the integrated charge. The collected charge is then obtained from the Most Probable Value (MPV) of a Landau-Gaussian convolution fitted to the distribution of the waveform integral. The uncertainty is taken as the difference between the MPVs obtained with and without the noise removal, and typically amounts to \SI{0.5}{\percent}.
 
\begin{figure}
    \centering
    \includegraphics[width=1\columnwidth]{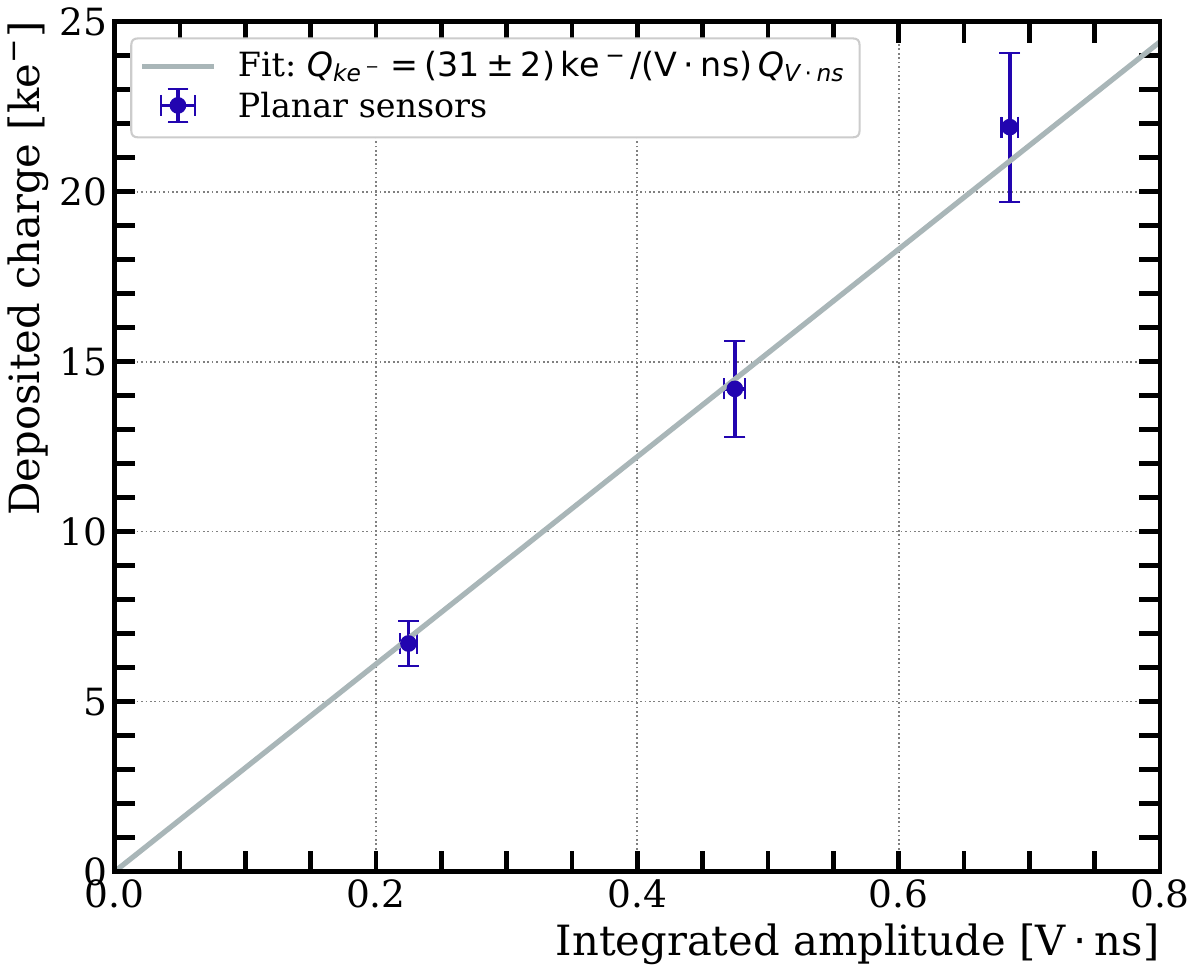}
    \caption{Charge expected from the PAI model as a function of the measured waveform integral, for the \SI{100}{\micro\meter}, \SI{200}{\micro\meter} and \SI{300}{\micro\meter} planar sensors. The line is a linear fit, whose slope gives the calibration constant of the OPTIMA channel.}
    \label{fig:planar_calibration}
\end{figure}

The charge expected in the three planar sensors is computed with the Photo Absorption Ionisation (PAI) model~\cite{PAI}, with an uncertainty of \SI{10}{\percent}~\cite{Riegler_2017}. Figure~\ref{fig:planar_calibration} shows it as a function of the measured waveform integral. The slope of a linear fit to these points gives the calibration constant between the waveform integral and the collected charge. This calibration was measured for each OPTIMA channel used for the \siemfive and \siemten data.
 
\subsection{Response of the SiEM as a function of \vgain}
The \siemfive and \siemten structures were placed in the DUT plane and their response to MIPs was measured at several values of \vgain, with \deltavbulk $= \SI{100}{\volt}$ to keep the bulk fully depleted. The gain voltage was varied from \SI{10}{\volt}, which is below the onset of multiplication, to the maximum value before breakdown of each structure. The collected charge was obtained from the waveform integral and converted to a number of electrons with the calibration of the corresponding OPTIMA channel. No selection is applied to the waveforms.
 
\begin{figure*}
    \centering
    \includegraphics[width=1\textwidth]{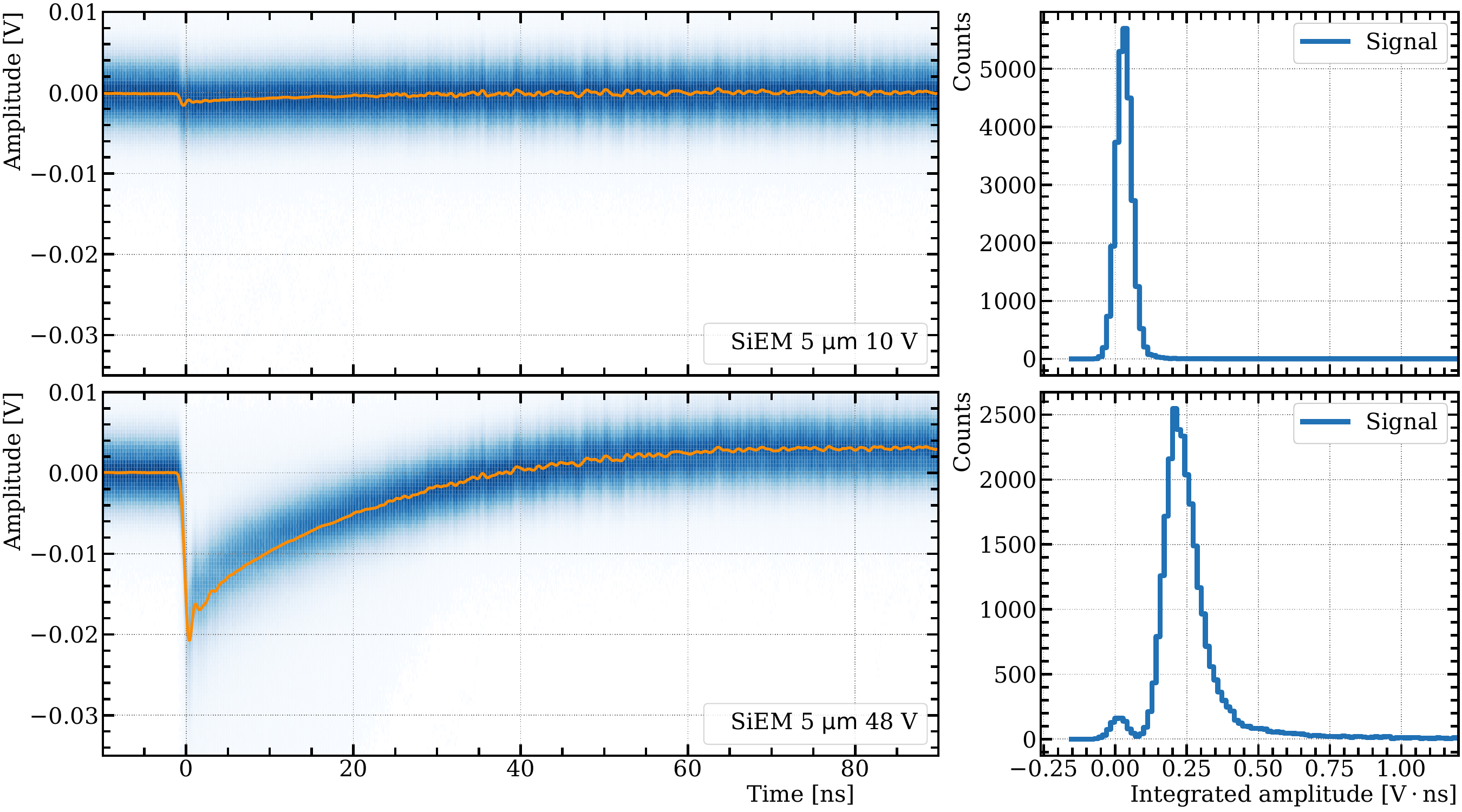}
    \caption{Left: waveforms recorded for a \siemfive structure at \vgain $= \SI{10}{\volt}$ (top) and \vgain $= \SI{48}{\volt}$ (bottom), with \deltavbulk $= \SI{100}{\volt}$. The colour scale indicates the number of events in each time-voltage bin and the orange curve is the average waveform. Right: corresponding distributions of the waveform integral. No selection is applied.}
    \label{fig:waveforms}
\end{figure*}

Figure~\ref{fig:waveforms} shows the waveforms recorded for a \siemfive structure at \vgain $= \SI{10}{\volt}$ and at \vgain $= \SI{48}{\volt}$. At \SI{10}{\volt} the signal is below the noise, and only the average waveform shows a clear response in time with the trigger. At \SI{48}{\volt} the signal is well above the noise, and the distribution of the waveform integral shows two well separated contributions, one from the signal in coincidence with the trigger and one compatible with noise.
 
The second contribution has multiple possible origins: a residual geometrical mismatch between the LGAD and SiEM active areas, non-parallel tracks crossing the ROI without passing through the SiEM, and triggers induced in the ROI pixel by particles crossing a neighbouring LGAD pixel. The first two are expected to be small. The latter is reduced by rejecting ROI waveforms whose shape is compatible with capacitive coupling rather than with direct charge collection. The remaining contribution is attributed to the intrinsic inefficiency of the structure, which the simulation of Section~\ref{sec:simulation} predicts for particles crossing the pixel far from the pillar.
 
Three methods are used to extract the response, depending on the size of the signal with respect to the noise. When the peak of the waveform-integral distribution is separated from the noise-integral distribution by more than five standard deviations, corresponding to \SI{0.12}{\volt.\nano\second}, the collected charge is obtained directly from the fit described above. The quoted uncertainty then combines the uncertainty of the charge calibration and the statistical uncertainty of the fit.

Below this separation, the collected charge cannot be extracted from the distribution of the waveform integral without assumptions on the transport efficiency and on the other contributions to the distribution. The signal shape is instead assumed to be independent of \vgain, with only its amplitude varying, and the collected charge is obtained by normalising the measured amplitude to a point where the charge is measured directly with the fit method. Two quantities are used to measure this amplitude: the number of waveforms above a given threshold on the waveform integral, and the maximum of the average waveform. Neither gives access to the collected charge directly, but both have a higher signal-to-background ratio than the distribution of the waveform integral. For the first, the uncertainty is statistical and reaches \SI{10}{\percent} of the measured value, as the method probes the tail of the distribution of the waveform integral. For the second, the uncertainty comes from the inefficiency contribution, which is unknown and may change the shape of the average waveform. Its effect is estimated from the ratio of the integral of the average waveform to its maximum, which is independent of the amplitude and therefore sensitive only to a change of shape. This ratio varies by up to \SI{20}{\percent} with respect to the normalisation point, and this variation is taken as the systematic uncertainty.
 
\begin{figure}
    \centering
    \includegraphics[width=1\columnwidth]{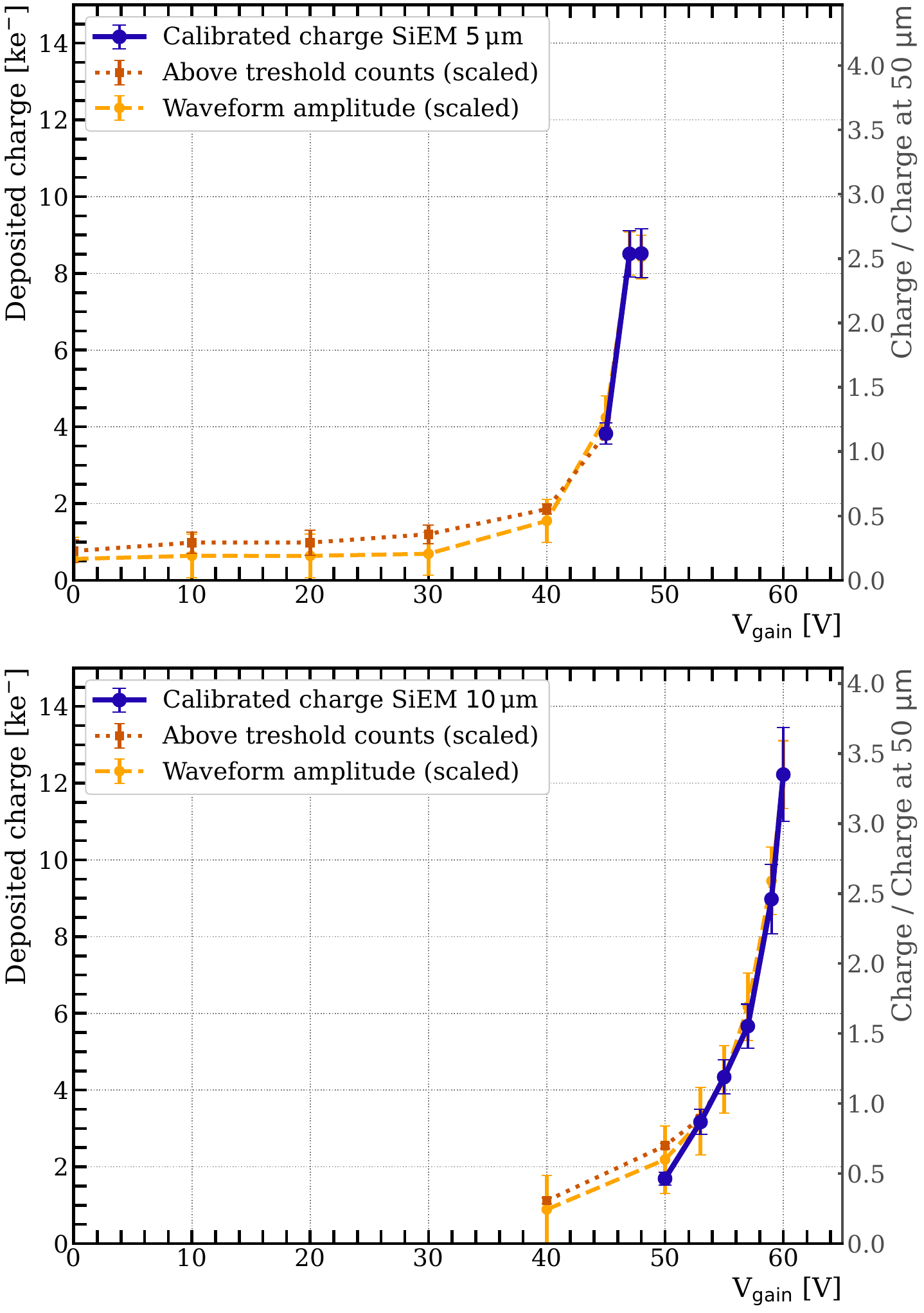}
    \caption{Collected charge as a function of \vgain for the \siemfive (top) and the \siemten (bottom), with \deltavbulk  $= \SI{100}{\volt}$. Blue: charge obtained from the fit to the distribution of the waveform integral and the calibration. Dark orange: charge obtained from the number of waveforms above a threshold. Light orange: charge obtained from the maximum of the average waveform. The right axis gives the charge normalised to that expected from a \SI{50}{\micro\meter} planar sensor without internal gain.}
    \label{fig:amplitude-vs-voltage}
\end{figure}

The collected charge obtained with the three methods is shown in Figure~\ref{fig:amplitude-vs-voltage} as a function of \vgain, for both structures.
 
\subsection{Evidence of charge multiplication}
The collected charge follows the behaviour observed in the I-V characteristics of Section~\ref{sec:electrical}: it is stable at low \vgain, and rises steeply above the onset voltage of each structure, which is lower for the \siemfive than for the \siemten. 
 
The SiEM structures have an active thickness of \SI{50}{\micro\meter}, so the collected charge can be compared with the charge expected from a \SI{50}{\micro\meter} planar sensor without internal gain. This comparison is given by the right-hand axis in the plots of Figure~\ref{fig:amplitude-vs-voltage}. The measured charge exceeds this reference by a factor of up to 3.5 at the highest accessible voltages. Since the charge collected without multiplication cannot exceed the charge deposited in the active thickness, a ratio above unity requires a gain larger than unity. The increase observed with \vgain is therefore not only due to an improvement of the charge transport, and the structures multiply the charge deposited by a MIP.
 
The simulation predicts a charge collection efficiency of about 5 for the \siemfive and about 6 for the \siemten at the pillar centre. The measured values are lower, which is expected: all pixels of a structure are interconnected and the position of the track in the pixel is not known, so the measurement averages the response over the whole pixel area, including the regions far from the pillar where the transport efficiency is low. 
 
\subsection{Time resolution}
A first measurement of the time resolution of the structures was performed with the same setup, using the LGAD of the first plane as the time reference. A resolution of about \SI{120}{\pico\second} is obtained for the \siemten. This value is dominated by the front-end electronics: all pixels of a structure are interconnected, so the capacitance seen by the OPTIMA input is large and the rise time of the signal is degraded. The contribution of the signal formation itself, in which the amplified electrons are collected almost immediately while the holes drift out of the pillar and across the bulk, cannot therefore be assessed with the present readout. Improving the resolution requires either connecting a smaller number of pixels together, or bonding the structures to an ASIC designed for fast timing, such as Timepix4. Both options are foreseen for the next HPK production, which will allow the intrinsic time resolution of this signal formation to be measured and compared with the simulation.

The dependence of the amplitude and of the time resolution on the temperature also remains to be measured. The $M_{10}$ curves of Section~\ref{sec:electrical} vary little between \SI{-20}{\celsius} and \SI{22}{\celsius}, which suggests that the gain, and therefore the signal amplitude, is largely independent of the temperature in this range. Cooling is nevertheless expected to increase both the ionisation coefficients and the carrier drift velocity at a given field~\cite{1677871}, and therefore to improve the
time resolution, at the cost of operating closer to the breakdown voltage. With a readout that is not limited by the front end, the next production could allow the amplitude and the time resolution to be measured as a function of the temperature, and this expectation to be tested.

\section{Conclusions and Outlook}
\label{sec:conclusions}
 
The first functional SiEM demonstrators, produced by HPK in a single-electrode configuration with lateral field injection, have been characterised electrically and with minimum ionising particles. Two structures were studied, the \siemfive and the \siemten, which differ in pitch and pillar geometry.
 
The TCAD and Garfield++ simulations of the two structures describe the depletion of the bulk, the onset of the multiplication in the pillar and the breakdown, as well as the evolution of the charge transport as a function of \vgain. 
The measured I-V characteristics reproduce the expected behaviour within about \SI{10}{\percent}.
 
The relative current increase $M_{10}$ is used as a proxy for the gain. Charge multiplication is reached at a lower \vgain for the \siemfive than for the \siemten, and is independent of the bulk voltage, between \SI{100}{\volt} and \SI{350}{\volt}, and of the temperature, between \SI{-20}{\celsius} and \SI{22}{\celsius}, up to the vicinity of the breakdown. This is the behaviour expected if the current increase is the multiplication of the thermally generated current, with a multiplication factor set by \vgain alone. 
 
The test beam measurements establish that the charge deposited by a MIP is multiplied as well. The collected charge follows the behaviour of $M_{10}$ below the breakdown region, and exceeds the charge expected from a \SI{50}{\micro\meter} planar sensor without internal gain by a factor of up to 3.5. Since the charge collected without multiplication cannot exceed the deposited charge, this demonstrates a gain larger than unity. The measured values are lower than the values predicted at the pillar centre, as expected for a measurement that averages over the whole pixel area. A first measurement of the time resolution gives about \SI{120}{\pico\second}, but at this stage is entirely dominated by the front-end electronics jitter.
 
These results establish that a gain defined by the geometry and the applied potentials, without a doped gain layer, can be obtained in a silicon sensor and applied to the charge deposited by a MIP. Several aspects remain to be studied.
 
The response is predicted to vary across the pixel, as the transport efficiency is lower far from the pillar. Data were taken with the same structures in a Timepix4 telescope~\cite{tpx4_telescope}, with the OPTIMA read out by a SAMPIC digitiser~\cite{sampic}. This setup allows the measured waveforms to be associated with the impact point of each track on the sensor.  The analysis of these data will allow the response to be measured as a function of the position in the pixel and compared with the simulated maps. The characterisation of the structure with laser, both single- and two-photon absorption setups, is also ongoing and will consolidate this understanding. Both will be the subject of a separate publication. 
 
The time resolution measurement is currently limited by the capacitance of the interconnected pixels. A next HPK production will include structures with a smaller number of interconnected pixels, and structures suited to the bonding of a fast timing ASIC, which will allow the achievable time resolution to be measured.
 
The radiation hardness, which motivates the concept, remains to be established. Samples have been irradiated at $10^{14}$, $10^{15}$, $10^{16}$ and $2\times10^{16}\mathrm{n_{eq}.cm^{-2}}$ and their characterisation is in progress.
 
\section*{Acknowledgements}
The authors thank Hamamatsu Photonics K.K. for taking on the development of the SiEM and producing the demonstrator structures. We thank the Timepix4 telescope team for the shared test beam time and for their help in cross-checking the alignment of the structures. We acknowledge the support of the SPS beam line coordinators for the test beam access, and of CERN EP-DT. The authors thank Florentina Manolescu and Marcin Poblocki from the CERN EP-DT bond lab, and Matthieu Heller and Gabriel Pelleriti from the University of Geneva, for their help in preparing the samples. This work was carried out in the context of the CERN Strategic R\&D Programme on Technologies for Future Experiments and received
funding from the European Union's Horizon 2020 Research and Innovation programme under grant agreement no.\ 101004761.


\section*{Declarations}
The authors declare no conflicts of interest. Data is available by contacting the corresponding author.

\section*{Abbreviations}
    The following abbreviations are used in this manuscript:
    \\
    \noindent
    \begin{tabular}{@{}ll}
        APD     & Avalanche Photodiode                  \\
        CCE     & Charge Collection Efficiency          \\
        CNM     & Centro Nacional de Microelectrónica   \\
        DRIE    & Deep Reactive Ion Etching             \\
        HPK     & Hamamatsu Photonics K.K.              \\
        I-V      & Current-Voltage                       \\
        LGAD    & Low Gain Avalanche Detector           \\
        MACEtch & Metal Assisted Chemical Etching       \\
        MEMS    & Micro-Electro-Mechanical Systems      \\
        MIP     & Minimum Ionising Particle             \\
        PAI     & Photo Absorption Ionisation           \\
        PSI     & Paul Scherrer Institute               \\
        ROI     & Region of Interest                    \\
        SiEM    & Silicon Electron Multiplier           \\
        TCAD    & Technology Computer-Aided Design
    \end{tabular}

\bibliographystyle{elsarticle-num}

\bibliography{main}

\end{document}